\documentstyle[twocolumn,12pt]{article}

\newcommand{\beq}{\begin{equation}}
\newcommand{\enq}{\end{equation}}
\newcommand{\cpl}{Chem. Phys. Lett.}
\newcommand{\jcp}{J. Chem. Phys.}
\newcommand{\jpc}{J. Phys. Chem.}
\newcommand{\prb}{Phys. Rev. B}
\newcommand{\prl}{Phys. Rev. Lett.}

\newcommand{\Si}[1]{\mbox{Si$_{#1}$}}

\begin{document}
\title{Chemistry of Nanoscale Semiconductor Clusters}
\author{Jun Pan,\thanks{Also, the Department of Physics,
New York University, New York, New York 10003-6621.}
Atul Bahel, and Mushti V. Ramakrishna \\*
The Department of Chemistry, New York University,
New York, NY 10003-6621.}
\date{}
\maketitle

\begin{abstract}

The ground state structures of small silicon clusters are determined
through exhaustive tight-binding molecular dynamics simulation
studies.  These simulations revealed that \Si{11} is an icosahedron
with one missing cap, \Si{12} is a complete icosahedron, \Si{13} is a
surface capped icosahedron, \Si{14} is a 4-4-4 layer structure with two
caps, \Si{15} is a 1-5-3-5-1 layer structure, and \Si{16} is a
partially closed cage consisting of five-membered rings.  The
characteristic feature of these clusters is that they are all surface.

Smalley and co-workers discovered that chemisorption reactivities of
silicon clusters vary over three orders of magnitude as a function of
cluster size.  In particular, they found that \Si{33}, \Si{39}, and
\Si{45} clusters are least reactive towards various reagents compared
to their immediate neighbors in size.  We provide insights into this
observed reactivity pattern through our stuffed fullerene model.  This
structural model consists of bulk-like core of five atoms surrounded by
fullerene-like surface.  Reconstruction of the ideal fullerene geometry
gives rise to four-fold coordinated crown atoms and $\pi$-bonded dimer
pairs.  This model yields unique structures for \Si{33}, \Si{39}, and
\Si{45} clusters without any dangling bonds and thus explains their
lowest reactivity towards chemisorption of closed shell reagents.  We
also explain why a) these clusters are substantially unreactive
compared to bulk surfaces and b) dissociative chemisorption occurs on
bulk surfaces while molecular chemisorption occurs on cluster
surfaces.  Finally, experiments on Si$_x$X$_y$ (X = B, Al, Ga, P, As,
AlP, GaAs) are suggested as a means of verifying the proposed model.

\end{abstract}

\section{Introduction}

Nanometer size particles are the embryonic forms of matter whose
microscopic study provides insights into the evolution of material
properties from molecules and surfaces to solids
\cite{Pool:90,Corcoran:90,Bjornholm:90,Siegel:93}.  Furthermore,
nanoscale particles have been shown to exhibit exotic optical
properties and reactivities quite different from those in molecules and
solids
\cite{Pool:90,Corcoran:90,Bjornholm:90,Siegel:93,Elkind:87,Jarrold:89}.
For these reasons, theoretical studies on clusters are critical to the
design and synthesis of advanced materials with desired optical,
electronic, and chemical properties.  Such studies are at the interface
of the traditional fields of quantum chemistry, solid state chemistry,
and statistical mechanics.  Hence, physicists, chemists, and material
scientists are working individually and in teams to unearth the
fundamental principles underlying the structure, dynamics, and
reactivities of these clusters
\cite{Bjornholm:90,Siegel:93,Elkind:87,Jarrold:89,Phillips:88,Jelski:88,Kaxiras:89,Bolding:90,Patterson:90,Swift:91,Roth:94}.
Indeed, the diverse group of scientists assembled at this {\sc
Nanomeeting-95} is testimonial to the breadth and depth of scientific
inquiry into this novel state of matter.

The nanoscale semiconductor clusters have potential applications in
non-linear optical devices, photovoltaic devices, and as photochemical
catalysts.  For this reason, intense experimental activity exists on
the synthesis, structural characterization, spectroscopy, and
reactivities  of these clusters.  On the other hand, the theoretical
and computational efforts on these systems are at a primitive stage.
In particular, due to both experimental and theoretical
limitations, the detailed knowledge of the role of surface structure on the
reactivities of these nanoparticles is lacking.  To correct
this imbalance, we started a systematic theoretical investigation of
the structure-reactivity relationship in semiconductor clusters.  The
results from these investigations is the subject of this Article.

\section{Structures of small silicon clusters}

Since naked silicon clusters are highly reactive, they are mostly
synthesized in a molecular beam under high vacuum conditions
\cite{Elkind:87,Jarrold:89}.  The number density of available clusters
is so low that diffraction based structural investigations are not
feasible under these experimental conditions.  Theoretical calculations
have now established the structures of \Si{N} clusters in the $N$ =
2-10 atom size range \cite{Krishnan:85,Rohlfing:90}.  These calculations have
employed
quantum chemistry molecular orbital techniques.  While these methods are
accurate for structural determination, computationally they are
highly cpu intensive and hence it is difficult to search all possible
cluster geometries.  Consequently, we employed the tight-binding
molecular dynamics (TB-MD) simulations for the determination of the
ground state
structures of \Si{N} clusters in the $N$ = 11-16 atom size range.  The
results of these simulations are presented below.

The choice of the tight-binding method for the study of the
cluster structures is motivated by its accuracy and computational
efficiency
\cite{Menon:91,Menon:93-1,Menon:93-2,Ordejon:94}.  This all valence
electron method is equivalent to the extended H\"{u}ckel method well
known in theoretical chemistry \cite{Lowe:78}.  This method has yielded
structures in excellent agreement with {\em ab initio} electronic
structure calculations for both carbon and silicon clusters
\cite{Menon:91,Menon:93-1,Menon:93-2,Ordejon:94,Tomanek:86}.
Recently, Menon and Subbaswamy have constructed an accurate
tight-binding Hamiltonian for silicon clusters
\cite{Menon:93-2,Ordejon:94}.  This Hamiltonian
includes Harrison's universal parameters appropriate for the
description of bulk Si \cite{Harrison:80}, supplemented with two to
four additional parameters for the description of the silicon
clusters.  These additional parameters are derived by fitting to the
\Si{2} bond length and vibrational frequency and to the overall
size-dependent cohesive energy curve of clusters
\cite{Menon:93-2,Ordejon:94}.  Most importantly, none of the parameters
of this Hamiltonian are fit to any of the cluster structures.
Full details of the Hamiltonian and computational
methods are described elsewhere
\cite{Menon:93-2,Ordejon:94}.

As the size of the cluster grows, the number of structural isomers
increases exponentially, with the result that searching the complete
configuration space for the global potential energy minimum becomes a
formidable task.  However, by combining the tight-binding method with
the molecular dynamics \cite{Allen:87} simulated annealing technique we
can efficiently search the cluster configuraton space and determine the
ground state geometry \cite{Bahel:95,Pan:95}.  We used this TB-MD
method in all the calculations reported here.

Figure 1 displays the lowest energy structures obtained using this
TB-MD technique for clusters in the $N$ = 11-16 atom size range.  The
\Si{11} structure consists of two tetragons in the
anti-prism geometry and three caps.  Two of these caps are attached to
the opposite faces of the two tetragons, while the third one is
attached to the edge of the top tetragon.  This structure may also be
described as an icosahedron with one missing atom.  Replacing this
missing atom would give rise to the icosahedral cage
structure for \Si{12}.  Such a spherical cage structure has not been
predicted or observed for any of the 12-atom elemental clusters.
Adding a face cap to \Si{12} gives the lowest energy \Si{13}
structure.  An alternative structure, derived by placing a Si atom
inside the cage of \Si{12}, is a high energy local minimum.

Adding a face cap to the \Si{13} structure does not yield the lowest
energy \Si{14} structure.  Instead, \Si{14} assumes a layer structure
consisting of three planes of four atoms each and two adjacent face
caps.  By suitably rotating this structure, we may also describe it as
a pentagon sandwich (or prism) with two caps each at the top and the
bottom.  The pentagon prism is somewhat distorted and displaced.  We
also considered a bi-capped hexagonal anti-prism as a candidate for the
ground state of \Si{14}.  However, this structure proved to be
unstable, indicating that six-atom ring structures are still not
favored in these small clusters.  Like \Si{14} cluster, \Si{15} assumes
a layer structure consisting of 1-5-3-5-1 layers.  However, \Si{16} is
quite unlike any of the previous clusters.  It is an open cage
consisting of fused pentagons, reminescent of the small fullerenes.

\section{Stuffed Fullerene Model}
Searching the complete configuration space for the ground state
structures is very difficult for
large clusters, even with the TB-MD method.  Consequently, we propose
a simple model that explains the trends in the reactivities of silicon
clusters in the 30-50 atom size regime.  We call this the stuffed
fullerene model \cite{RK:94}.  This model consists of 1) a central atom
A, 2) four atoms B surrounding A in tetrahedral geometry, and 3)
fullerene surface.  The B atoms bind to twelve surface atoms, thus
rendering the B atoms also bulk-like with four-fold coordination and
tetrahedral geometry.  The surface then relaxes from its ideal
fullerene geometry, the same way the $(1 \times 1)$ bulk surfaces
relax.  This relaxation gives rise to crown atoms and dimers (CAD)
pattern on the surface.  The crown atoms are formally three-fold
coordinated and possess one dangling bond each.  The dimers are also
formally three-fold coordinated, but they eliminate their dangling
bonds through $\pi$ bonding.  The essential feature of this
construction is that the bulk-like core of five atoms (A + B) and
fullerene-like surface make these structures stable.  In principle,
this model is applicable to clusters containing more than twenty
atoms.

Unlike carbon, silicon does not form strong delocalized $\pi$ bonds
\cite{Brenner:91}.  Consequently, fullerene cage structures
\cite{Curl:91,Boo:92,Fowler:92} are energetically unfavorable for
silicon clusters \cite{Menon:94}.  Instead, intermediate sized silicon
clusters prefer $\sigma$-bonded network structures similar to the
diamond structure of bulk silicon \cite{Menon:94}.  The fullerene
geometry for the surface, consisting of interlocking pentagons and
hexagons, gives special stability to the surface atoms \cite{Boo:92}.
Furthermore, since delocalized $\pi$ bonding is not favorable in
silicon, we expect the surface atoms to relax from their ideal
fullerene geometry to allow for dimer formation through strong local
$\pi$ bonding.  Our model accounts for all these facts.

We generate the \Si{33} structure by stuffing the 5-atom core inside
the \Si{28} fullerene cage.
We orient the 5-atom pyramid in such a way that the
central atom A, the core atom B, and the crown atom C lie on a line.
The crown atom is at the center of three fused pentagons and it is
surrounded by three other surface atoms D.  The D atoms now relax
inwards to form the B-D bond.  The same type of relaxation motion is
necessary to form the $2 \times 1$ reconstruction on the bulk Si(111)
surface \cite{Cohen:84,Lannoo:91}.  The activation barrier of $\approx$
0.01 eV \cite{Cohen:84,Northrup:82} for this relaxation is easily
recovered by the formation of the B-D bond, whose strength is 2.3
eV/bond \cite{Brenner:91,Lannoo:91}.  Consequently, such a relaxation
of fullerene surface is feasible even at 100 K.  Finally, the remaining
surface atoms E readjust to form as many dimers as possible.  Similar
relaxation motion and dimer formation occurs on Si(100) surfaces also
\cite{Lannoo:91,Chadi:79}.  The dimers are $\sigma$-bonded pair of
atoms whose dangling bonds are saturated through the formation of $\pi$
bonds.

The structures of \Si{33}, \Si{39}, and \Si{45} clusters thus generated
are displayed in Fig. 2.  These structures reveal that the crown atoms
are able to form a fourth bond to the core atoms B, thus rendering the
B atoms formally five-fold coordinated.  The B-C bond arises from the
back donation of the electrons from C to B and it weakens the
neighboring bonds through electronic repulsion.  We do not know the
strength of this back bond, but it is sufficiently strong to eliminate
the dangling bond on the crown atom and make these magic number
clusters unreactive.

Chemisorption is sensitive to the local electronic structure and
chemical bonding on the surface.  On bulk Si(111)-($7 \times 7$)
surface, chemisorption occurs preferentially on rest atom sites and
less on crown atom sites \cite{Wolkow:88}.  Absence of rest atoms in
silicon clusters is the main reason why these clusters are much less
reactive than bulk silicon.  Furthermore, dissociative chemisorption of
ammonia on bulk Si(111)-($7 \times 7$) surface is due to the close
proximity of the rest atom and crown atom sites \cite{Wolkow:88}.
Absence of such a configuration on silicon clusters explains why
reagents are unable to dissociate on the cluster surfaces; instead they
chemisorb molecularly.   We predict that the alloy clusters
Si$_x$X$_y$ (X = B, Al, Ga, P, As, AlP, GaAs) will be highly inert and
that it may be possible to synthesize these in macroscopic quantities.  We
suggest experiments on these alloy clusters as a means of fully
exploring the role of surface structure on reactivities.

\section{Summary}

In summary, we determined the structures of small silicon clusters
through TB-MD simulations.  These simulations revealed that \Si{11} is
an incomplete icosahedron, \Si{12} is a complete icosahedron, \Si{13}
is a surface capped icosahedron, \Si{14} is a 4-4-4 layer structure,
\Si{15} is a 1-5-3--5-1 layer structure, and \Si{16} is a partially
closed cage consisting of fused pentagons, reminescent of small
fullerenes.    The atoms in all these clusters strongly prefer to lie
on the surface rather than inside.

We also propose a novel structural model to explain the experimental
data of Smalley and co-workers on the chemisorption reactivities of
silicon clusters towards various reagents \cite{Elkind:87}.  This model
consists of bulk-like core of five atoms stuffed inside reconstructed
fullerene cages.  The resulting structures of \Si{33}, \Si{39}, and
\Si{45} are unique, have maximum number of four-fold coordinated atoms,
minimum number of surface atoms, and zero dangling bonds.  This model
does not yield such unique structures for other intermediate sized
clusters and hence they will have larger number of dangling bonds.
This explains why \Si{33}, \Si{39}, and \Si{45} clusters are least
reactive towards reagents with closed shell electronic structure, such
as ammonia, methanol, ethylene, and water \cite{Elkind:87}.  We suggest
that experiments on alloy clusters are needed to fully understand the
role of surface structure on the chemical reactivities of these
nanometer size particles.

\section*{Acknowledgments}

This is the fifth paper in this series on {\em Chemistry of
Semiconductor Clusters}.  This research is supported by the New York
University Research Challenge Fund and the Donors of The Petroleum
Research Fund (ACS-PRF \# 26488-G), administered by the American
Chemical Society.  All the computations reported in this paper were
carried out on an IBM 580 Workstation.  The graphics presented here are
generated using the XMol program from the Research Equipment Inc. and
the University of Minnesota Supercomputer Center.

\newpage

\begin{figure}
\caption{The lowest energy structures of \Si{N} ($N$ = 11-16)
clusters, determined using the tight-binding molecular dynamics
simulated annealing technique.}
\end{figure}

\begin{figure}
\caption{Structures of the least reactive  clusters, \Si{33},
\Si{39}, and \Si{45}, obtained using the proposed stuffed fullerene
model.   These clusters do not possess any dangling bonds and hence are
least reactive towards reagents with closed shell electronic structure,
such as ammonia, methanol, ethylene, and water.  Representative atoms
in different chemical environments are labeled from A to E.  }
\end{figure}

\end{document}